\documentclass[a4paper]{PoS}

\title{Approach to the time development of parton fragmentation}

\ShortTitle{Approach to the time development of parton fragmentation}

\author{\speaker{Martin Rohrmoser}
\\SUBATECH, 
4 rue Alfred Kastler, 
44072 Nantes, Cedex 03, France\\
        E-mail: \email{martin.rohrmoser@subatech.in2p3.fr}}

\author{Pol-Bernard Gossiaux\\
        SUBATECH, 
        4 rue Alfred Kastler, 44072 Nantes, Cedex 03, France\\
        E-mail: \email{gossiaux@subatech.in2p3.fr}}
        
\author{Thierry Gousset\\
        SUBATECH, 
        4 rue Alfred Kastler, 44072 Nantes, Cedex 03, France\\
        E-mail: \email{gousset@subatech.in2p3.fr}}
        
\author{J{\"o}rg Aichelin\\
        SUBATECH, 
        4 rue Alfred Kastler, 44072 Nantes, Cedex 03, France\\
        E-mail: \email{aichelin@subatech.in2p3.fr}}
\abstract{It is the central goal of our studies to describe parton fragmentation in the hot and dense medium of a quark gluon plasma (QGP). Under the assumption that the medium is not static and homogeneous, knowledge about the temporal evolution of the processes involved can be of essential importance. Therefore, parton fragmentation has been studied with a Monte-Carlo algorithm that approximates the DGLAP-evolution of fragmentation functions via a set of parton cascades. The presented work consists mainly of implementations of this kind of algorithm and the application of a simple approximation, which gives the time development of partonic cascades.  
As approximations to the parton-splitting processes of gluons and massless quarks in the vacuum two different schemes with the same leading-log contributions are investigated. Furthermore, the temporal evolutions of quantities related to partonic cascades have been examined: To our understanding, especially variables (like, e.g.: parton multiplicities or virtualities) that give insight, when parton splittings take place, are of great interest, as they allow identifying the parts of the fragmentation process that will be most affected by medium interactions.}

\FullConference{The European Physical Society Conference on High Energy Physics\\
		22-29 July 2015\\
		Vienna, Austria}

\begin{document}

\section{Introduction}
Parton showers that are produced in heavy-ion collisions and propagate in the hot and dense medium of a QGP can be simulated in event generators that rely on Monte-Carlo algorithms~\cite{herwig,pythia,sherpa}.
For partons propagating in the vacuum, the perturbative part of their fragmentation process can be approximated via Monte-Carlo simulations (cf., e.g.,~\cite{Renk,Zapp}, which contain also treatments for the fragmentation in a medium.) of the DGLAP evolution~\cite{Dokshitser,AP}, 
\begin{equation}
\sigma_{n+1}=
\sigma_n\otimes \frac{\alpha_s}{2\pi}P_{a\rightarrow b+c}\,,\label{eq:dglap}
\end{equation}
where the total cross section $\sigma_{n+1}$ factorises into a hard part $\sigma_n$ and splitting functions $P_{a\rightarrow b+c}$ for the splitting of parton $a$ into partons $b$ and $c$. 
In this kind of algorithms, the parton splitting in vacuum is mainly described by its longitudinal momentum fraction, which follows the distribution of the splitting functions, together with the parton virtuality $Q$, which is selected between an infrared cut-off $Q_0$ and an upper boundary $Q_{max}$ from the partition function of the Sudakov factor ${S}_a\left(Q_{max},Q\right)$  (cf.~\cite{ESW} for a review),
\begin{equation}
S_a\left(Q_{max},Q\right)=\exp\left(-\int_{Q^2}^{Q_{max}^2}\frac{d\tilde{Q}^2}{\tilde{Q}^2}\int_{x_-}^{x_+}dx\frac{\alpha_s(x,\tilde{Q}^2)}{2\pi}\sum_{b,c}P_{a\rightarrow b+c}(x)\right)\,,\label{eq:sudakov}
\end{equation}
where $x$ is defined as a momentum fraction of parton $b$ as compared to parton $a$. The boundaries $x_\pm$ are defined in Sec.~\ref{sec:s1vss2}. Soft emissions are resummed by the depency of $\alpha_s$ on $x$.
From Eq.~(\ref{eq:sudakov}) it follows that parton splitting in the vacuum can be entirely described in the momentum space.

In the medium, partonic cascades evolve in a time dependent and spatially non-homogeneous environment -- expressed by a changing temperature $T$ and other medium properties. As the medium interactions, as well as their probabilities, largely depend on $T$, it is necessary to employ a space-time dependent picture for the evolution of the partonic cascade. 

The present work systematically studies the temporal development of a cascade in the vacuum, which is initiated by a single light quark with fixed total energy $E_1$. In particular, the number $N(t)$ of particles produced, their virtualities $Q(t)$, and furthermore their spatial distributions are investigated. Finally, it is evaluated, how these quantities are affected by different corrections to the leading-log approximation to parton splitting.
\section{Spatio-temporal picture of partonic cascades}
A frequently used approximation~\cite{Renk,Zappdiss} (applicable only to a probabilistic approach of parton fragmentation, as the one used here) for the time $\Delta t$ between production and decay of a virtual parton is given by 
\begin{equation}
\Delta t=\frac{E}{Q^2}\,,\label{eq:lifetime}
\end{equation}
where $E$ is the parton energy in the laboratory frame.
In order to obtain the spatial points of parton splittings, it is assumed that, between $2$ successive splittings, the intermediate partons behave as free partons in addition to Eq.~(\ref{eq:lifetime}). Subsequently, the following steps $\Delta r^\mu$ separate the creation and decay of a virtual parton in Minkowski space:
\begin{equation}
\Delta r^\mu=\frac{p^\mu}{E}\frac{E}{Q^2}=\frac{p^\mu}{Q^2}.
\end {equation}
The spatial distribution of partons in a cascade can, thus, be simulated at any given time $t$.
\section{Approaches to partonic cascades}\label{sec:s1vss2}
At leading-log accuracy the QCD radiation-pattern is captured with leading-order splitting-functions by the DGLAP evolution, Eq.~(\ref{eq:dglap}). This approximation is applicable, if $Q_i\ll E_i$ and $Q_b,\,Q_c\ll Q_a$ for parton energies $E_i$ and virtualities $Q_i$ ($i=a,\,b,\,c$). For a first instance of a Monte-Carlo algorithm (here, referred to as scheme 1) one considers that the fraction $x$ in Eq.~(\ref{eq:sudakov}) is a light-cone-energy fraction, $z$ (cf.~\cite{ESW}). For a splitting 
\begin{figure}
\centering
\includegraphics[scale=0.85]{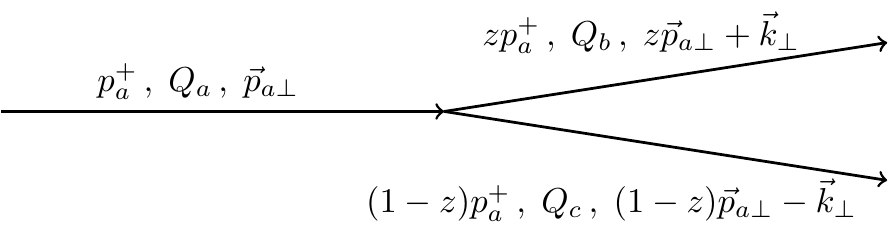}
\end{figure}

it follows that
\begin{equation}
\vec{k}_\perp^2=z(1-z)Q_a^2-(1-z)Q_b^2-zQ_c^2\,.\label{eq:kt}
\end{equation}
It is imposed that $Q_a^2\geq Q_b^2,\,Q_c^2\geq Q_0^2$ and $|\vec{k}_\perp|\geq f \Lambda_{QCD}$ (with a numerical constant $f\approx 1$; this article takes $f=1.1$ from~\cite{Zappdiss}.) for a splitting to be accepted. For the corresponding integration boundaries in $z$ one obtains
$
z_\pm=\frac{1}{2}\left(1\pm\sqrt{1-\frac{4Q_0^2+4(f\Lambda_{QCD})^2}{Q_a^2}}\right)
$. 
Since no reference frame needs to be specified, the evolution of the shower is frame independent in this formulation.

In an alternative approach (here, referred to as scheme 2) the fraction $x$ can be taken as an energy fraction~\cite{Zapp,Zappdiss} with the integration boundaries 
$
x_\pm=\frac{1}{2}\left(1\pm\sqrt{\left(1-\frac{4Q_0^2+4(f\Lambda_{QCD})^2}{Q_a^2}\right)\left(1-\frac{Q_a^2}{E_a^2}\right)}\right)
$. 
Thus, the splittings are more likely to be rejected, especially, when $E_a$ is not much larger than $Q_a$. The occurrence of energy in the rejection criterion makes scheme 2 frame dependent.
\begin{figure}[b]
\hbox{\hspace{-0.3cm}
\includegraphics[scale=0.99]{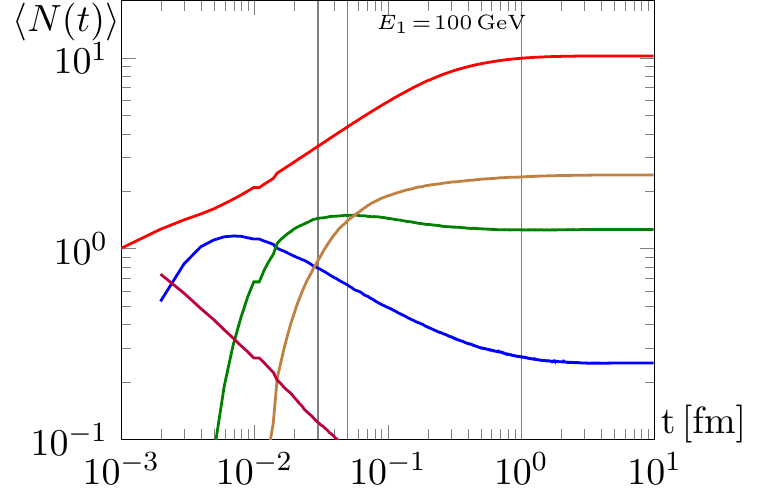}
\hspace{0.4cm}
\includegraphics[scale=0.99]{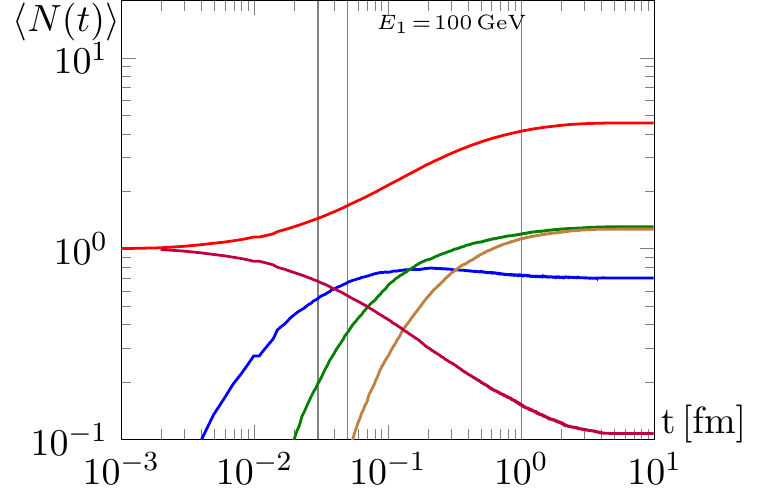}\\}
\hbox{\hspace{-1cm}
\includegraphics[scale=0.99]{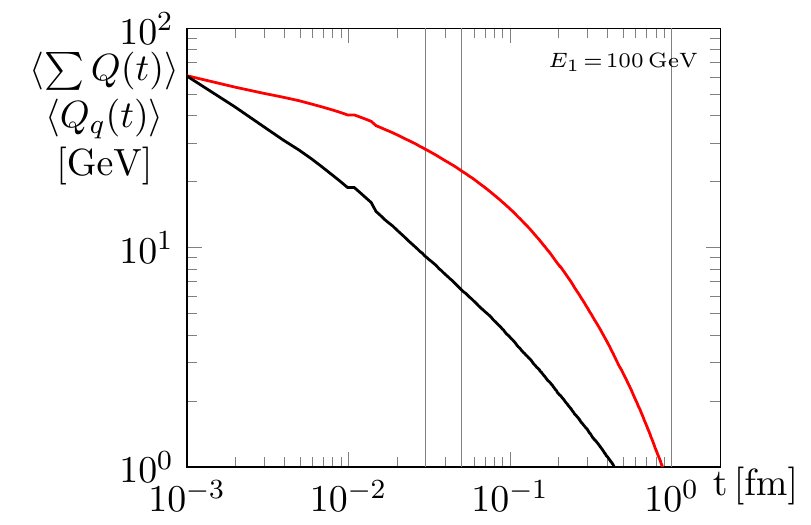}
\includegraphics[scale=0.99]{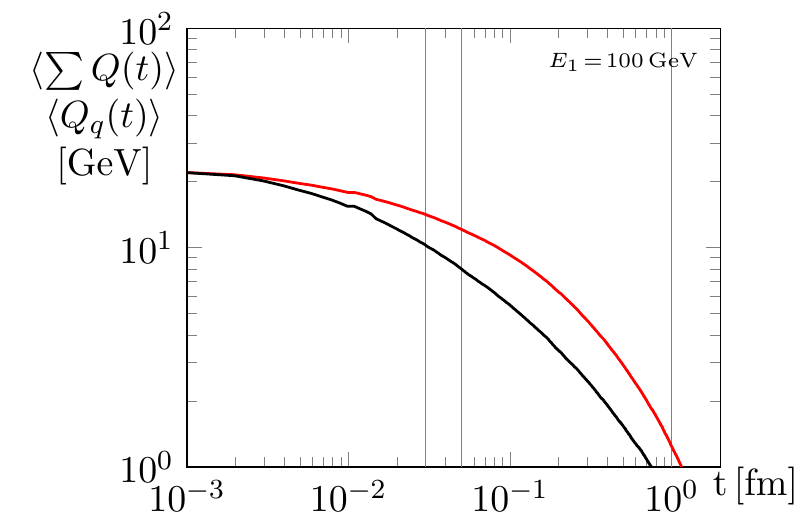}}
\caption{Upper panels: The average number $\langle N(t)\rangle$ of partons present in a partonic shower at time $t$ (red), together with its contributions from partons produced in $0$ (purple), $1$ (blue), $2$ (green), and $3$ (brown) successive splittings. Lower panels: The average total virtuality $\langle\sum Q(t)\rangle$ per cascade (red) together with the average virtuality $\langle Q_q(t)\rangle$ of a tagged initial quark (black) as a function of time $t$. Left side: results from scheme 1. Right side: scheme 2. }\label{fig:NtQt}
\end{figure}
\begin{figure}[b]
\includegraphics[scale=.49,clip=true,trim=0cm 0.1cm 0cm 0.cm]
{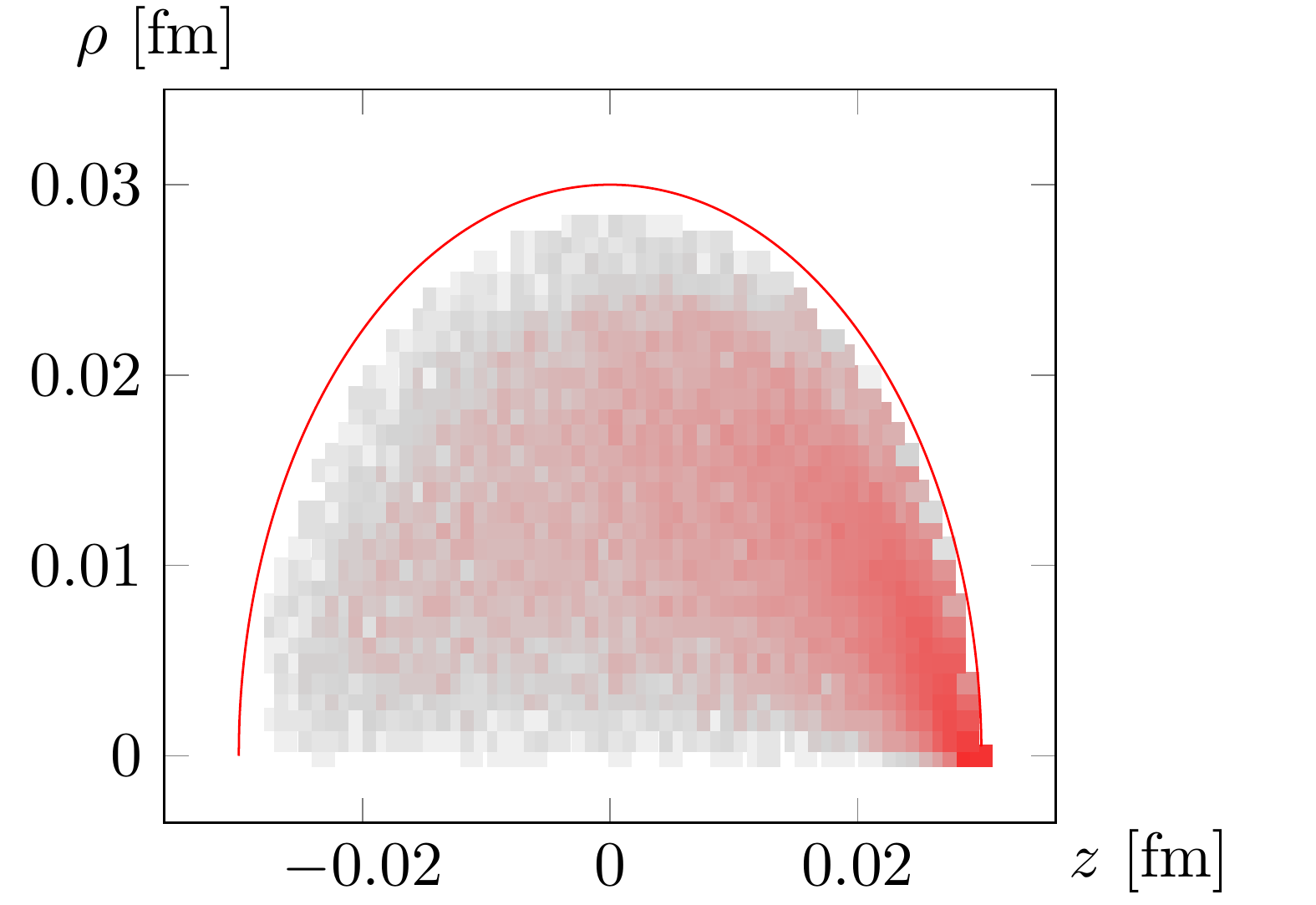}
\includegraphics[scale=.49,clip=true,trim=0.0cm 0.1cm 0cm 0.0cm]
{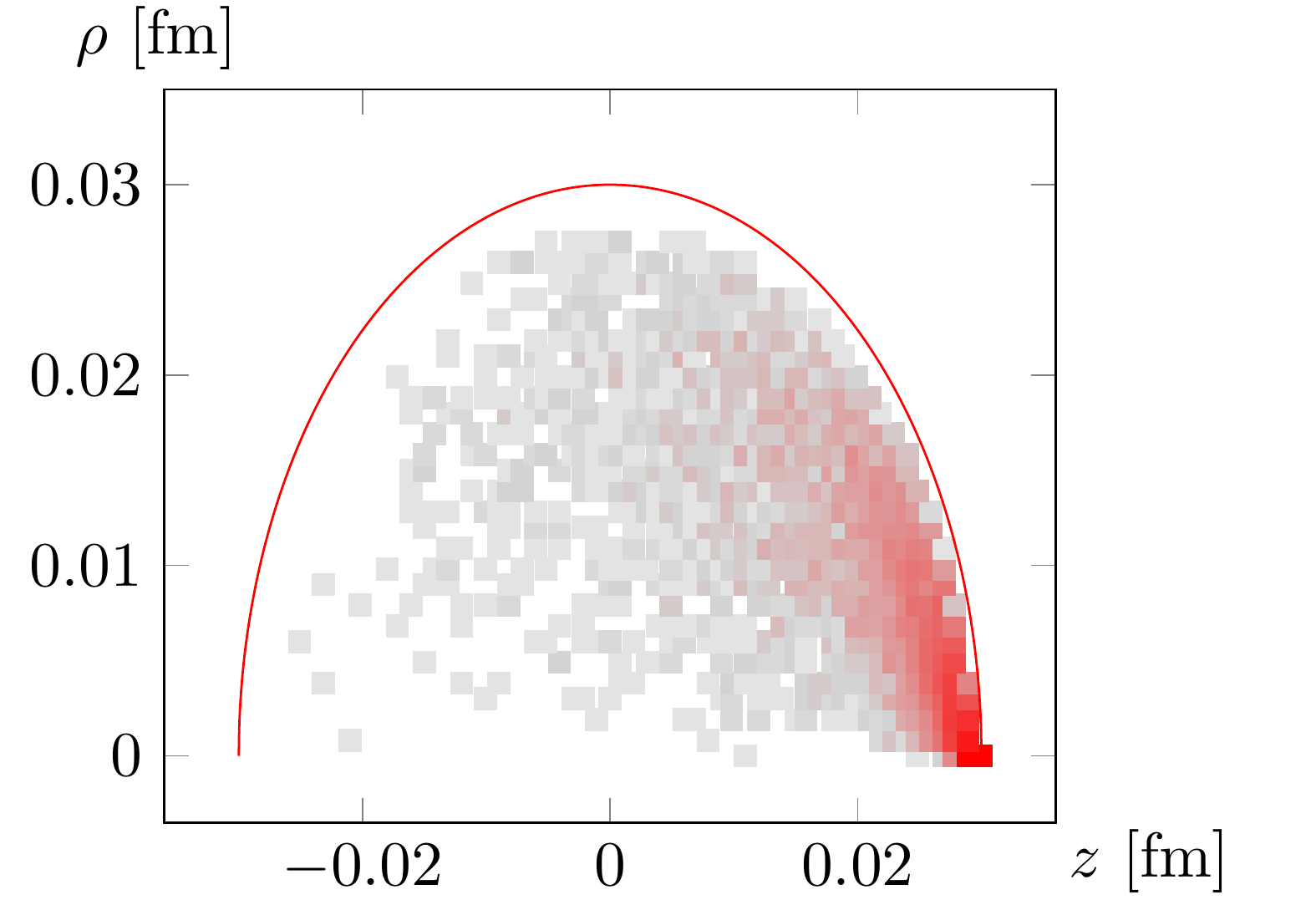}\\[-0.1cm]
\includegraphics[scale=.49,clip=true,trim=0cm 0.1cm 0cm 0.cm]{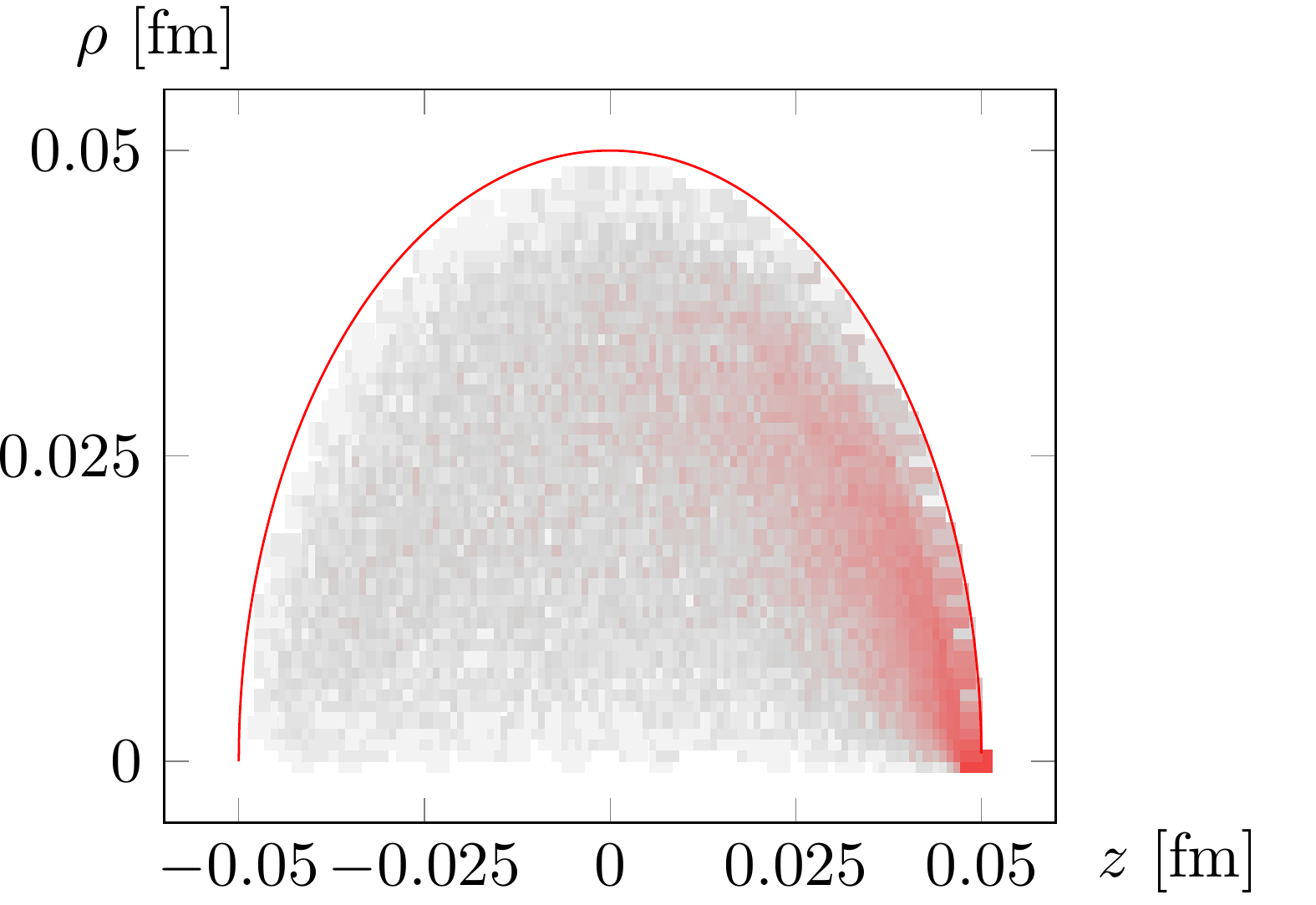}
\includegraphics[scale=.49,clip=true,trim=0cm 0.1cm 0cm 0.cm]{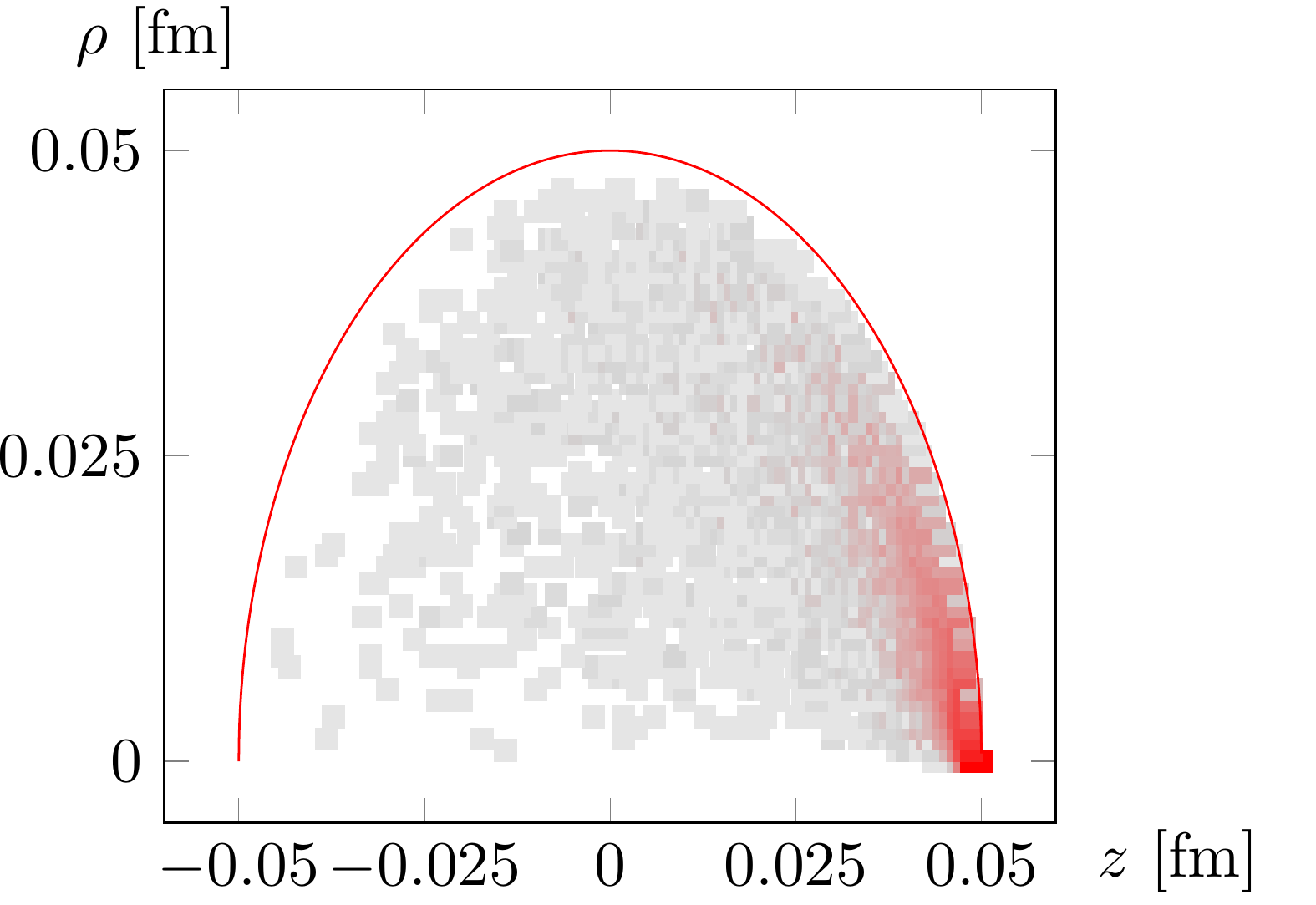}\quad\\
\includegraphics[scale=.49,clip=true,trim=0.cm 0.cm 4.9cm 0cm]
{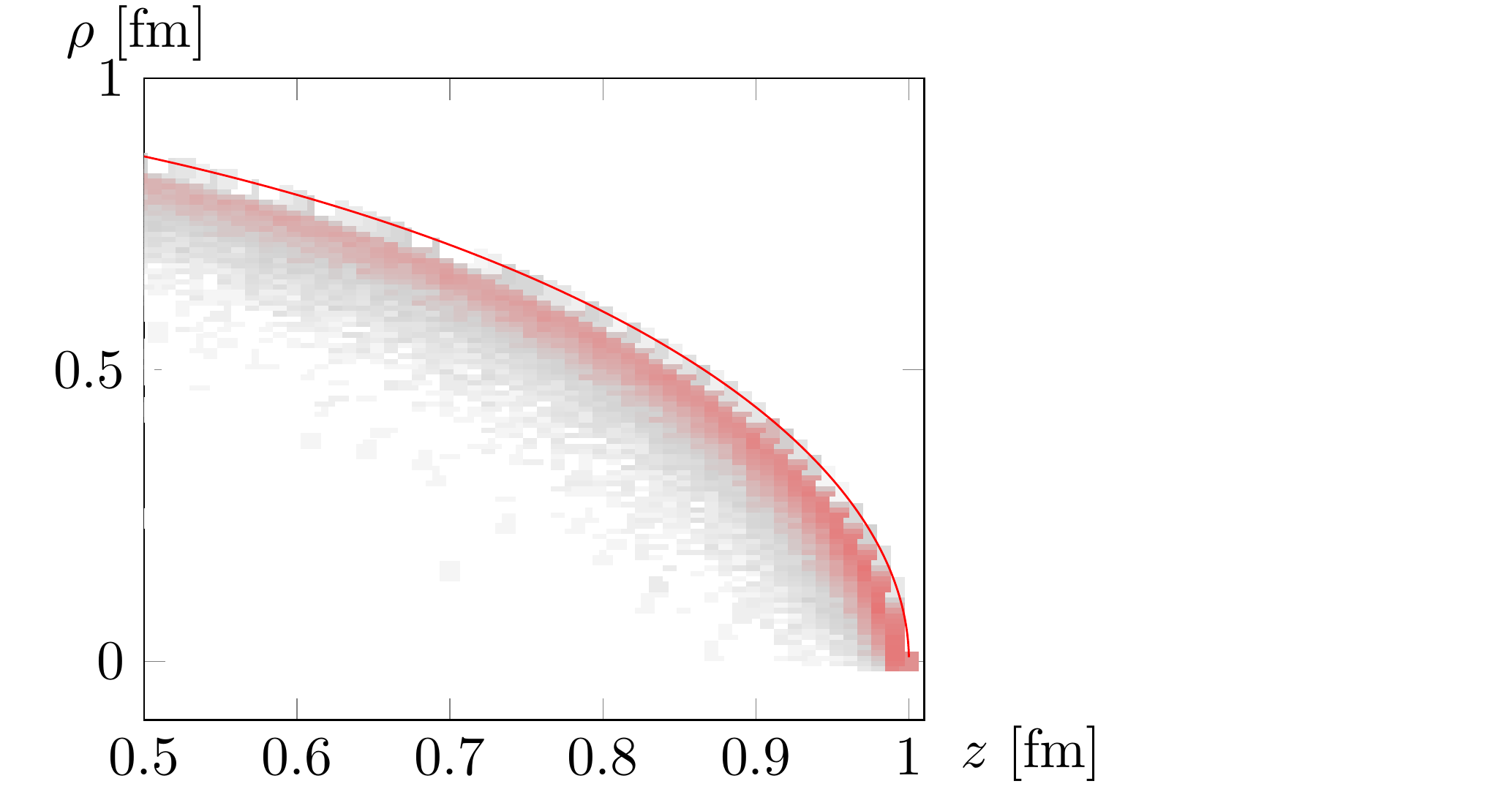}
\includegraphics[scale=.49,clip=true,trim=0.0cm 8.cm 4.9cm 0cm]{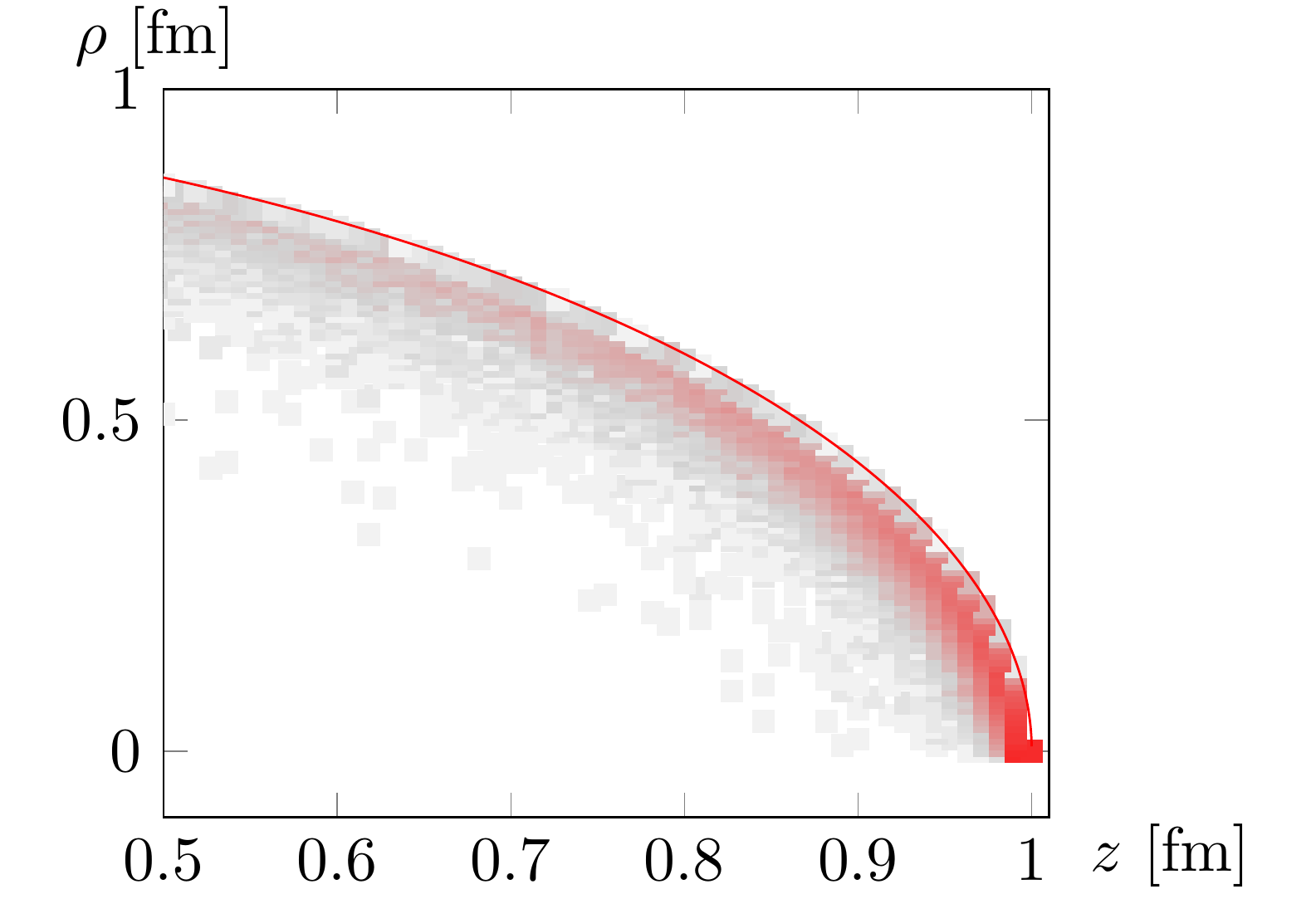}\\
\includegraphics[scale=.49]{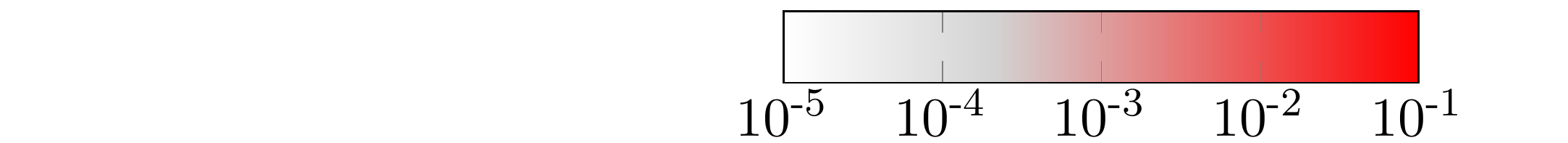}
\caption{Parton-number density per pixel in longitudinal direction $z$ and transverse directions $\rho=\sqrt{x^2+y^2}$ at the times $t=0.03$~fm (upper panels), $t=0.05$~fm (middle panels), and $1$~fm (lower panels). In the diagrams above, a pixel corresponds to a square with the side length $\frac{t}{110}$. Left side: results from scheme 1. Right side: scheme 2. Particles travelling with velocities smaller than the speed of light have to be found in the space constrained by the red circles.}\label{fig:galettes}
\end{figure}
\section{Results}
In order to simulate particle showers in the vacuum, a Monte-Carlo simulation of the DGLAP evolution of partons, based on the parton evolution in vacuum of~\cite{Zapp} was implemented. The necessary input for this program is the energy $E_1$ and an upper limit $Q_{max}$ for the virtuality of the initial parton in the cascade. At the threshold of $Q_0=1$~GeV it is assumed that no further parton splittings occur and that the partons hadronize. As an example, $e^+$-$e^-$ collisions in the center-of-momentum frame that produce an outgoing quark-antiquark pair were assumed at a scale of $\sqrt{s}=200$~GeV. The spatial and temporal evolution of one of the outgoing jets was investigated, under the assumption that $Q_{max}=E_1=\frac{\sqrt{s}}{2}$: The mean number of partons of a cascade at time $t$, $\left\langle N(t)\right\rangle$ is shown in the upper panels of Fig.~\ref{fig:NtQt} for both schemes. The mean cascade saturates with $10$ partons per cascade at $t\approx1$~fm for scheme 1 and with $5$ partons per cascade at $t\approx3$~fm for scheme 2.
The numbers of partons that were either initially present, or produced via $1$, $2$, or $3$ successive splittings are also shown in Fig.~\ref{fig:NtQt} for both schemes. For scheme 1 the mean number of initially present quarks decreases like a power law with time at early times. Later on, the asymptotic value is reached. Since in the subsequent generations particles are produced, while others decay into the next generation at the same time, a plateau is formed, where creation and decay are at equilibrium. After a time of $t\approx1$~fm for scheme 1 or $t\approx3$~fm for scheme 2, respectively, a threshold of particles, which does not decay further, is still present. It can be observed, again, that the time scales in scheme 2 are of a factor approximately $3$ larger than in scheme 1, while a smaller numbers of partons is produced. 

The activity of the partonic cascades was also investigated by the corresponding virtualities: The sum of the virtualities of all partons per cascade, averaged over the total number of simulated cascades, is shown in the lower panels of Fig.~\ref{fig:NtQt}, together with the averaged virtuality of a single (leading) parton, here the initial quark. Both curves indicate that the vacuum evolution of the cascades stops at the order of magnitude of $1$~fm, since the threshold of  $Q_0=1$~GeV is already attained before. Most remarkably, for scheme 2 the virtualities at small time scales are far smaller than for scheme 1.

Furthermore, the spatial distribution of the simulated partons was studied. Some sample results can be seen in Fig.~\ref{fig:galettes}: A Cartesian coordinate system $(x,y,z)$ with the $3$-momentum of the initial quark of the cascade parallel to the $z$-axis was assumed. Fig.~\ref{fig:galettes} shows the parton-number density per pixel as a function of $\rho=\sqrt{x^2+y^2}$ and $z$ for $t=0.03$~fm, $0.05$~fm, and $1$~fm. These time scales are also indicated (in grey) in Fig.~\ref{fig:NtQt}. In order to  compensate for the expansion with time, the pixel size is rescaled accordingly. Also, since the number of partons increases with time, the density is normalised to $1$. The light front, $\rho^2+z^2=t^2$, is also shown in Fig.~\ref{fig:galettes}. 
The parton-number density exhibits a considerable peak in forward direction which becomes increasingly pronounced at larger time scales, with a relatively small spatial dispersion in comparison to earlier times. This peak is more strongly collimated in the results obtained from scheme 2, than in those from scheme 1. Thus, the average parton shower that is simulated via scheme 1 can be assumed to probe a larger volume, when entering a hypothetical medium, than the average cascades obtained from scheme 2.
\section{Conclusions}
Monte-Carlo simulations of the DGLAP evolution of a light quark in the vacuum, have been implemented via 2 different schemes that are the same up to leading-log accuracy, but differ in their phase-space constraints. To the resulting partonic cascades a spatio-temporal picture has been applied. For the 2 schemes, it has been found for initial quark energies, $E_1=100$~GeV that most partonic cascades stop in their evolution before a time scale of $t\approx1$~fm, or $t\approx3$~fm, respectively. In comparison to the time scales of the QGP medium this kind of behaviour would imply that most splittings happen shortly after the formation of the medium, or even before and corresponding observables are created on the same time scales. 
Furthermore, a picture of the spatial evolution of partonic cascades was used to show that most of the partons are concentrated in the forward direction of the shower, with rather small dispersion, especially at large times. 

For both schemes we found qualitatively a similar behaviour: However, in scheme 2, due to the different constraints in phase space (leading to a Sudakov factor peaked at small virtualities), parton splittings occurs at lower virtualities $Q$ and, thus, also at larger time scales.

\end{document}